\documentclass[12pt]{article}
\begin{document}

\baselineskip=5.0mm

\begin{center}

{\bf Yuval Ne'eman and How He Influenced My Life and Career}\\

\vspace{.25in}

Shmuel Nussinov\\
{\small email: nussinov@post.tau.ac.il}

Tel Aviv University, Sackler School Faculty of Sciences, \\
Tel Aviv 69978, Israel\\ and\\ Schmid Science Center Chapman University,\\Orange, California (USA)
\vspace{.12in}

{\small \it (July 2005)}

\vspace{.25in}

{\bf Foreword}
\end{center}

These notes written four years ago shortly after Yuval Ne'eman's eightieth
birthday, describe a few events that happened mainly during 1961-1964.

  These are very specific events pertinent to my scientific education, to
  some specific aspects of early theoretical high energy physics in Israel
  connected with the late Yuval Ne'man and the small group that worked
  around him at that time. Since I was away during the equally
  important period of 1964-1968 and missed the dramatic pre-string
  developments and various plans for new high energy theory centers,
  these notes do not pretend to present any---albeit approximate---history of
  theoretical high energy physics in Israel.   In particular I do not
   address the important impact that Yuval had also on experimental high energy physics.
 
  It was because of my full realization of these shortcomings that I
  refrained from posting these notes. I realize, though, that while reflecting
  a strong personal bias some of these may be of interest to those who care
  not only about the final finished product, but also about the torturous
  ways leading to them.
 
 About three years ago, Yuval passed away. Prior to that I was in close contact
 with him regarding the nice biography ``Soldier, Scientist and Statesman"
 written on him by Andrew Watson. I have finally decided on the occasion of
 my own seventieth birthday to post these notes.

\begin{center}
* * * * * * * * * * * * * * *
\end{center}

  In the summer of 1961, I was 22 years old. I and three members of my class had just finished our physics Master's thesis at the Weizmann Institute.  We needed special
  permissions to do so since only the Hebrew University in Jerusalem (HUJI) and the Technion in Haifa could grant degrees at that time.

  As often the case I was a latecomer. Hearing that Joe,
  Haim and Arnon were going to work with Igal Talmi and Amos de-Shalit at
   Weizmann on theoretical nuclear physics, I wanted to do the same.

  As I entered the Physics building in Weizmann I asked the first
  person I met there, who turned out to be Prof. Gideon Yekutieli,
  where the offices of Talmi and deShalit were. Gideon, who was a
   particle physics experimenter, told me about the interesting high
   energy collisions in cosmic rays that he was investigating and
   convinced me to work with him instead.

  The four of us were good students: the impressive, witty Haim
  Harari (Yuval's first cousin), the exuberant Arnon Dreiman-Dar,
 Joe Ditkovsky  (Yossef Dothan), the most scholarly
  among us, and myself.

   Dothan was the best student in our class graduating ``Summa cum
   Laude".  I might have been close but missed a simple question on
   the final oral exam with Racah, Cohen and Rechavy. The argumentative
   Arnon almost blew it telling Rechavy that he does not understand
   some basic physics, and Harari's brilliant performance there made
   up for years of minimal study.

  At that time I heard of an Israeli intelligence colonel,
  Yuval Ne'eman, who in two years finished a PhD with Abdus Salam at
  Imperial College and whose SU(3) group was a true revolution.\footnote{SU(3)
  was used by Sakata and his school in Japan. They however used the
  fundamental, triplet representation for the known P (proton) N (neutron)
  nucleons and the $\Lambda$ ($\lambda$) hyperon. The idea being that all other
  hadrons are composed of these and/or their anti-particles. (Indeed, in
  the early days of the quark model we and many others used p, n and
  $\lambda$ instead of the
  (u) up, (d) down and (s) strange quarks which {\it do} comprise the
  fundamental representation.) The all-important novelty of Gell-Mann -
  Ne'eman's ``Eight-fold Way" was using the adjoint (octet)
  representation for all known baryons.}
 I was deeply impressed and wanted to work with him.

  When Ne'eman returned shortly thereafter to the Israeli AEC (Atomic Energy
  Commission), he tried to continue his SU(3) research.  To this
  he was joined by Asher Gotsman, his peer from Imperial, Haim Goldberg,
  Racah's ex-assistant, David Horn from the Technion, Yuval's MSc and
  later PhD student, and by Joe, Haim Harari, and myself.

Relativity and other theoretical research work was being done at the
   Technion, Yuval's Israeli Alma Mater, where he graduated much earlier
   in electrical and mechanical engineering. Yet research in nuclear physics
   and group theory, the disciplines closest to Yuval's recent SU(3) work, was
   done mainly at the Weizmann Institute and the Hebrew University.

  A recognized authority in group theory, Giulio Racah, was the senior
  professor at HUJI. I recall his heavy Italian accent in the courses on
  electromagnetism, quantum  mechanics and group theory that we had with
   him and admired his ability to lecture with no notes (and vary from year to year!).

    At the Institute of Advanced Studies (IAS), Princeton, I met in 1974-76
  Tullio Regge from the Regge poles which were so crucial in my carrier.
  He truly admired Racah and was particularly proud of proving new relations
   for the ``Racah Coefficients" and of convincing the community to name a
   lunar valley after Racah.
   Inspired by Majoranna's work, Racah suggested early on a process
   akin to neutrino-less double beta decay---a ``Holy Grail" of much
   research to this very day.
   He gave at the IAS a series of lectures on Lie algebras/groups which may
   have been attended  by the young Murray Gell-Mann.

    It would seem natural for Racah to espouse and promote Yuval the young
   Israeli star in the field that was his life-long passion.

   Racah lectured on SU(3) in a 1962  Istanbul summer school but did not
  directly help Yuval's research in Israel. I was told that at his untimely
  death in 1965 the thesis of Haim Harari
  on SU(3) of which Racah was a referee, was lying in his drawer.
  At that time Racah  was HUJE's president and Yuval
   became one of the founding fathers of the competing Tel Aviv University (TAU).

  The Nuclear physics theoreticians Zvi (Harry) Lipkin, Sydney Meshkov and Carl
   Levinson from Weizmann were inspired by Yuval and started most intensely
   and successfully to apply SU(3). Indeed they were quite familiar with
   another version where O(3) rather than SU(2) was naturally embedded in SU(3).

   In contrast I believe that initially our (Joe, Haim H., and myself) own
   efforts to help Yuval were not very successful.

   Indeed our physics education at HUJI was lopsided and we knew nothing about
   scattering  and field theory, a fact that I fully realized only
   when attending Carl Levinson's course at Weizmann and some lectures there by
   S. Schweber.

   What Ne'eman needed at that time were experienced researchers steeped
   in particle phenomenology and knowledgeable in the above areas.

   I wonder if had he opted not to return right away after his PhD to the
   demanding post in the AEC, but rather fully devoted himself to
  developing and applying his SU(3), things might have been different...

  In just two years at Imperial College he closed huge gaps in his
  physics education and learned enough to realize that SU(3) in its
   Eight-fold version provided the correct classification.
  Two extra years of intense research with access to data might have led
   him to full-fledged quarks and to sharing the Nobel Prize with Gell-Mann.

  However, after four years abroad he and his family may have been homesick.

  Also the defense ministry having given him two precious years to study
   for his thesis at Imperial College may have wanted him back in service.

  Yuval might have hoped for a triumphant return and immediate acceptance
  in the Israeli academy. But, to the best of my knowledge, no Israeli
   institute offered him {\it right away} a professorship and an opportunity
   to build a new group around him.

  Untempered by previous failures, Yuval may have been confident that he
   could be an administrator in the AEC and lead the research on SU(3).

  Was Yuval thinking of Leibnitz, the ambassador, or of Carnot, the
   colonel, the co-inventors of calculus and thermodynamics?

  There was a bit of these and many other physics and math greats in him.
  His associative memory was legendary. He had remarkable curiosity and
  knew more about almost everything than anyone else. He could weave a
   rich and (coherent!) yarn of tales about Greek and Jewish sages, about
   royalties, historic warfares, about evolution of creatures and ideas and
   about almost everything under the sun. Watching this ``force of nature",
   this rare and unique human was an awe-inspiring experience.

   Yuval was very busy and could not see us during most of the time.
  We would get short notes from him describing ideas he had during
  administrative duties and meetings or while driving around in between.

  In one note he suggested that I look at a paper by Gell-Mann and
  Levy on ``sigma models". I studied it at length, but lacking in
   background could not really understand it.
   As often the case, Yuval's remarkable intuition was correct. A bit
  later, PCAC, most clearly introduced in this paper, became a powerful
   and useful tool--the precursor of effective chiral lagrangians.

  Yuval's powers of concentration were incredible and stayed with him for
   many years. I recall that about two decades later while we were both
   visiting the IAS, Dvora,  Yuval's wife, became seriously ill and was
   hospitalized for several weeks. The devoted Yuval never left her
   bedside---yet in the interim managed to write a physics paper.

   Our physics discussions with Ne'eman were often interrupted by phone calls.
   After awhile a pause occurred while the other party was looking for something,
  Yuval would then turn to us, pick up the discussion exactly at the point we
  stopped and made a new observation that occurred to him while talking on the phone.

  This style, reminiscent of simultaneous chess games, is adequate
  when playing with amateurs but not with the world class Gell-Mann.
  In the 1960's Gell-Mann dominated theoretical particle physics
  overshadowing anyone else to an extent which is hard to imagine
  (except, perhaps, by string theorists, thinking of Ed Witten).
  This 31 year old prodigy from Caltech---several years Yuval's junior in
   age---had an amazing record already as early as 1960:

  He introduced strangeness (independently with Nishijima);
  explained K-$\bar{K}$ oscillations with Pais; suggested the
  renormalization group with Low and the V-A (Conserved Vector
  Current) theory for weak interactions with Feynman. 

   Gell-Mann did not have Yuval's remarkable extracurricular career.
   He did not fight in real wars, contribute critically to his country's
    security, initiate new universities, new political parties or channels
    connecting different seas. Yet he knew every bit of physics worth knowing
    such as the existence of an esoteric cosmic ray event seen by Yehuda
    Eisenberg which looked like an $\Omega^-$ production and
    decay.\footnote{Another revealing incident to which I can testify, took place in the
Fermi Lab conference in 1972 in Chicago---the fateful period when most members of the Israeli Olympic team in Munich were murdered by terrorists.
  I was sitting by myself relatively early in the morning in the hotel's lobby
  watching the news. I was then joined by Gell-Mann with whom I was slightly
  acquainted. Knowing that I may be familiar with Lenny Susskind, a
  co-inventer of strings, who visited and deeply influenced our group at TAU,
  Gell-Mann then addressed me in something like the following:
  ``Is Mr.~``Sweet-Child's(??)" string nothing but Schwinger's line integral
  $(\int A_{mu}dx^{mu})$ along a path connecting a quark and an anti-quark??"
  I do not recall what I said in response. While according to Lenny, ``Suskind"
  originated from a Spanish ``Susskin" and not the German ``Suse/Zis-Kind"), this comment
is extremely insightful physics.  It connects the strings- originally motivated
  by the duality of resonances and Regge poles and the Veneziano model with the
  field theoretic idea that Kenneth Wilson suggested two years later as the
  underlying reason and mechanism for quark confinement.
  Gell-Mann was very much acquainted with and hedging his bets, actively worked
  on both the S matrix and the field theoretic approaches.}

  Feynman was the charismatic person at Caltech with whom Gell-Mann
  fiercely competed. Still, in particle physics proper he had no equal.

  The remarkable coincidence of him simultaneously suggesting the
  same SU(3) classification scheme helped Yuval initially.

  It was not clear that SU(3) or any other scheme based on any group
  was correct. The fact that also Gell-Mann, too, suggested the Eight-fold" SU(3) prompted
  experiments looking for the missing particles predicted and verifying
   that these, as well as some of the known particles, have the correct
   spins and parities so as to fit within the multiplets to which they
   were assigned. A great scientific drama was unfolding: will the many
   new particles/resonances fit?, and will the triply-strange ``$\Omega^-$
   be found?

  The relative rates of different 2 particles $\rightarrow$ 2 particles, if dominated by intermediate states in the s channel in specific SU(3) representations is fixed, in the
SU(3) symmetry limit,  just by group theoretical Clebsch-Gordan (C.G.) coefficients.
Yuval, Joe and Haims G. and H. analyzed in this way various reactions of protons and anti-protons. To generalize this to 2 $\rightarrow$ 3 processes Joe and Haim computed-in what seemed to me at the time a heroic effort- relevant C.G. coefficients,

The Weizmann group of Levinson, Lipkin and Meshkov found that most SU(3) predictions follow from just U(``You") and V (``We") spin analogs of I-(iso)spin.

  In a small side project with G. Yekutieli and H. Goldberg we found deviations from
SU(3) predictions due to  splittings of ``Regge Poles", the trajectories in the angular momentum, mass$^2$ plane.  These entities introduced by G.Chew and S.Frautschi following Regge's analytic continuation of angular momenta, contain many higher spin recurrences of known particles and dictated the high energy behavior of various 2 particle $\rightarrow$ 2 particle processes.

   And then in 1964 the $\Omega^-$ was discovered at Brookhaven National Lab (BNL) while Yuval was visiting Gell-Mann at Caltech for a year.
 
   The impact of this discovery at that time was monumental. It indicated the power of symmetry considerations which turned out to be---at least in this case---far superior to detailed dynamical calculations. Except for the positron and anti-proton, it was the first in a series of particles predicted jointly by symmetry and dynamics- such as the W and Z bosons, the charm c quark, the $\nu_{\tau}$ and the bottom b and the top t quarks.

  For a long time the BNL team (which included Yona Oren, later at TAU) searched for the telltale decay of the $\Omega^-$ in bubble chamber pictures with no success.

  It was also pointed out that dynamical effects due to the nearby kaon and the cascade baryon $K\Xi$ threshold may drastically shift the predicted value of the mass of $\Omega^-$.

As a joke, Gell-Mann was said to have gone to Japan at that time so that should the results from BNL be negative he could jump off mount Fuji...

  Shortly after the discovery both were interviewed. Yuval was remarking on
  how fortunate he was to work with Salam rather than with Bondi (on general
  relativity a subject that Yuval really loved) and that this was due to
  the geography of London: the long travel time from the Israeli Embassy
  to Kings College (where Bondi was) as compared to the much shorter commute
  to Salam's Imperial College.
  I have seen Yuval on many occasions after and before and he was never happier.
  It seemed like he felt that he will share the glory and the Nobel prize. In Israel
  he became a national hero and many popular articles explained to the layman the
   importance of $\Omega^-$.

   However Gell-Mann got most of the credit and in 1969 a Nobel Prize all by
  himself. The detailed report of the Nobel Committee cited Gell-Mann's
  strangeness, the quark model ``Of great heuristic value" and current algebra. As for
  the all-important ``Eight-fold" way, it was mentioned that it was conceived also by
  Yuval Ne'eman but his role as minimized.
  It would have been really nice if Yuval and Murray would have shared the prize given to
   a ``dream team" of a brilliant professional and equally brilliant colonel turned physicist.

  Yet the Nobel Committee's decision was rightly influenced by the most relevant fact
  that Gell-Mann (and also Sossumo Okubo from Rochester) found a relation between masses in SU(3) multiplets embodying a simple pattern of symmetry breaking which Gell-Mann used to predict the $\Omega^-$ mass with amazing precision.

 The quark model proposed by Gell-Mann and independently by George Zweig explained SU(3)and was the precursor to the present day QCD--Quantum Chromodynamics. The latter, dynamical theory, treats hadrons in the same way that QED, Quantum Electrodynamics, treats atoms.

  A puzzling feature, namely, the difficulty of reconciling the naive model in
  which the low-lying baryons are three quarks in S wave with spins coupled
  to 1/2 and 3/2 (in Gell-Mann--Ne'eman's octet and decuplet representations,
   respectively)
  led O.~W.~Greenberg to suggest parastatistics which was equivalent to introducing
  an extra internal degree of freedom which M.~Y.~Han and Y.~Nambu suggested
  to be gauged. In hindsight this is simply color.

   The formal abstract version of Gell-Mann's quark model led to current algebra which, in turn, made J.~D.~Bjorken predict ``scaling". Completing a full circuit, this and the experimental discovery of scaling in deep inelastic electron scattering by Nobel prize winners  J.~Freedman, H.~Kendall and R.~Taylor, led back to point-like quarks inside the proton.

  Finally the puzzling almost free behavior of the quarks at short distances
  was explained by asymptotic freedom, the remarkable feature of QCD, and
  non-abelian gauge theories in general, and for which a recent Nobel Prize
  in particle physics was awarded to D.~Gross, D.~Politzer and
  F.~Wilczeck.

    As in many other cases, Yuval also anticipated the quark model.
    He and Haim Goldberg introduced in 1962---a full two years ahead of Gell-Mann
    and Zweig---entities carrying 1/3 baryon number belonging in the fundamental
    triplet representation of SU(3). Since baryons are then made of three B=1/3
    entities, the direct product $3 \times 3 \times 3=10+8+8+1$ picks the eight-fold,
    ten-fold and singlet SU(3) representations for the baryons seen in nature.
  Unfortunately this simple fact is not mentioned in the rather formal paper
  which was not reprinted in the ``Eight-fold way" book by Gell-Mann and Ne'man that came
  out in 1964.

  Yuval went on to impressive military, academic and political careers nicely described in Yuval's biography that Andrew Watson wrote. In addition to Horn, Harari, Dothan who were to varying formal degrees his students, Yuval had many others including Y.~Achiman, M.~Gronau and A.~Aharony. They and their students, Yuval's spiritual ``grandchildren", constitute a major almost dominant part of Israeli high energy physicists in Israel and abroad.

   Yuval founded the Physics Department at Tel Aviv hiring  dozens of physicists in
  the space of two to three years in many areas including experimental high energy,
  condensed matter and astronomy and was deeply involved in building the
  Wise observatory in the Negev.

He played a key role in helping ``Refuseniks"---Jewish scientists who refused to collaborate with the regime, expressed their desire to leave the Soviet Union, and were fired from their universities. Yuval granted them positions at TAU even while in Russia and conducted famous joint seminars by telephone. It may well in part be due to these efforts that many prominent ex-Soviet physicists and mathematicians came to TAU.

   Over and above all this the fascination with fundamental physics that Yuval's success instilled in young Israelis led many of them to theoretical physics in
  which Israel still excels.

   Let me return then to our initial group and time.
   After finishing their PhD's, Joe, David and Haim H. went on as postdocs to
   Caltech and to Stanford and were all remarkably successful.
   I believe that Yuval's friendly and less authoritative attitude contributed
   to their independence and this success.

   In his thesis with Yuval and Lipkin, Harari finally ``proved" the eight-fold
   SU(3) by using  better experimental data and calculational techniques.
   At Weizmann, which he joined after SLAC, Haim invented duality diagrams in parallel with J.~Rosner (then a postdoc at TAU) and a two-component  Regge + Pomeron theory for high energy collisions.
  He suggested the ``Rishon" model---a constituent model for quarks and leptons.
  Unfortunately his rather elegant model suffered from ``anomalies" and could not
  provide a consistent field theory.
  After further work on quark and neutrino mixings he served for a  very long
  time as a successful president of the Weizmann institute.

  In the thesis D.~Horn did with Yuval he applied SU(3) to weak interactions
  missing the ``Cabibbo theory" of weak interactions since one piece of
  experimental information (eventually proven wrong) conflicted with it.
  During his postdoc at Caltech David worked with Gell-Mann. With C. Schmid, a Swiss post doc there, he invented ``finite energy sum rules". The ``Horn-Schmidt duality" that eventually emerged from those became a forerunner
  of the Veneziano model and the all-important string theory. In TAU he worked
  on hadronic phenomenology, lattice gauge theory, and on the connection between
  statistical mechanics and field theory, and recently on neural nets and bio-informatics.

   At Caltech Joe collaborated with Adler, Dashen, Gell-Mann and with Yuval
   who was visiting there in 1965. He was fascinated by the exact algebraic
  treatment of the Hydrogen atom and he and Yuval widely extended this as
  part of an effort to algebraize everything by introducing dynamical groups
  to particle physics. At TAU Joe was involved, in particular, in inventing
  geometric Fermions with Banks and Horn and kept learning and teaching
  mathematical physics in voluntary special courses, preparing the many Israeli physicists who attended those lectures for the eventual forward leap in
  practicing the mathematically involved string theory. Sadly Joe passed
  away 20 years ago.

  Lipkin continued working on SU(3). Inspired by the 3:2 ratio of proton-proton and pion-proton cross sections noted by Frankfurt and Levin then at
  the Soviet Union (now at TAU) he espoused very early on the quark model.

  He worked on these issues with Florian~Scheck.  With Gideon Alexander he made the important observation re the power of Zweig's ``No hair-pin quark diagram" rule in high energy reactions. He suggested the EPR-like correlation which introduces an important time structure in $\bar{b}-b$ decays.

He greatly promoted  the constituent quark model becoming  a famous and prolific researcher in this field in which he is active to this very day at the age of 88!.

   In his thesis work with A.~De-Shalit, A.~Dar invented diffractive production
  with absorptive corrections in the nuclear context.  Together with Kugler and Dothan, the four of us extended it to particle physics a subject pushed much further soon thereafter by Gottfried and Jackson. He joined the Technion at the time the rest of us joined TAU and Weizmann. Later he switched to astroparticle physics making original important suggestions in connection with neutrinos from supernovae.

He remained as argumentative as ever, strongly promoting at present together with A.~De-Rujula a model where emission of ``cannon balls" from supernovae is the source of both Gamma ray bursts and high energy cosmic rays.

 Carl Levinson became effectively my PhD mentor after I worked with his brilliant student, Moshe Kugler, on Regge poles. However, Carl fell in love
  with a young tour guide at Weizmann. Not being sure wether he wanted to continue in physics, he suggested that I continue my thesis with Ivan Muzinich,
  his collaborator in Seattle. After finishing
    my thesis, I had a two-year postdoc in Princeton, missing the excitements of
    SU(6), and early dual models that the other young Israelis enjoyed. In 1968
    I came back to TAU which Joe and David already joined. I did not work on the
    quark model/QCD until the $J/\psi$ charm quark composite was discovered
    in 1974, barely in time to get a piece of the action.

   After writing the influential Eight-fold Way review with Gell-Mann, Yuval
    continued physics research in between his many other jobs. He suggested
    algebraic schemes for Regge residues with Nicola Cabibbo and Lawrence Horowitz .
    Far more daring was the suggestion with Jan Thiery Mieg  predicting
    the top, W/Z and Higgs boson masses using ``Graded" Lie algebras; a theme
    that he kept coming back to, e.g., in a review he wrote with D.~Fairlie and
    S.~Sternberg.  Yuval collaborated with Gell-Mann and with Regge on supergravity
    and supersymmetry. He made the profound observation that a flat metric can be a ``Higgs-like"
    phase of a more general theory of space-time. He extended the Cartan
    classification of Lie algebras and representations to a new infinite spinorial representation.
   He also suggested a lagging core model which effectively is a multi-big-bang
   scenario.

   He also wrote books on the algebraic approach to particle physics and
   on strings and membranes long before these became a hot topic. He wrote
   popular books on particle physics, on evolution and on the inflationary
   universe.

\begin{center}
* * * * * * * * * * * * * * *
\end{center}

  It was a tiring Monday with the Yuval eightieth birthday fest
  lectures in the morning and afternoon at TAU, a long dinner at the
  Israeli Academy in Jerusalem, and many after-dinner talks lauding
  Yuval's many-fold talents and contributions. I kept watching Yuval
    sitting at the next table amidst his close family.

   He surely enjoyed most of it, but his head kept drooping. When finally he
   got up and hesitantly made his way towards the microphone I sensed some
   tension. Would Yuval live up to this moment? Or would age, tiredness and
the Parkinson's disease that he had for some time get the better of him?

  Yuval had a slow start--mentioned the influence of his aunt, the math
  teacher, and Haim's mother, and reminisced about an early discovery of
  compounded operations yielding gigantic numbers which in his typical
  prophetic style, he termed ``Googleplexes".

  But then he picked up and I could feel the good old Yuval shining through.
  He recalled an amusing incident where a newsman wrote after an interview
  that Yuval had academic, military, civil and aeronautical (?) careers. The
  puzzle finally resolved when he recalled being asked,``How do you manage
  to do so many things?" to which he responded: ``I work also on planes..."

  He closed by recounting meeting Churchill's daughter who told him
    about an amusing interview Churchill had on his 80'th birthday.
    In answering a question about a water level-like mark on the lower parts
  of the walls in the big hall at 10 Downing St., Churchill said that it is due
 to the champagne spilled at each of his previous birthdays and, pointing towards
  the ceiling he concluded, ``and so much more to go."

   After the long applause some one at our table noted that on such
   occasions the honored person usually thanks colleagues, students and
   institutions. The only people Yuval thanked there or in the conclusion
   the next day were his wife and his devoted secretary, Matilda.

  Realizing that this hardly surprised me motivated these notes and
  the following closing apologetica.

  The curious and remarkable fact was that while so many of us benefited
   from Yuval's knowledge, wisdom and influence there were very few, like the
   late Abdus Salam, whom  he should have been thankful to...

  The Israeli academia did not espouse him early on offering him a university
   position right away helping him to proceed with his promising research.

  Whoever pressured him to accept the AEC position had, in effect, robbed him
   of the chance he had to go for the gold when the time was right. Also
    notwithstanding our honest desire, at first we did not push forward much
    his SU(3) research.

  With QCD's hindsight, Gell-Mann--Ne'eman's ``flavor" SU(3) is an accident due
    to the ``up", ``down" and ``strange" quarks having the same color coupling
    and similar masses.

  Yet SU(3) played a crucial role comparable to the Mendeleev periodic table.
   Both put some order in the jumble of elements/particles classifying them in
   nice groups and predicting new elements/particles with specific prescribed
  properties just so as to fill ``gaps".

  Also in both cases an underlying composite structure (nucleus and  quantum
mechanical electrons--or quarks) and binding dynamics (QED or QCD), eventually
  emerged. However, the explanation of the periodic table was found only after
  half a century whereas the quark  model came in less than four years after
   flavor SU(3).

 
  There is a famous Hebrew saying which I'll distort to ``never put yourself
   in other's position". Ignoring it let me imagine myself in Yuval's place
   which is highly presumptuous. Independent of any formal recognition, the
   knowledge that I have seen the light that others failed to see, that nature
   allowed me to place a few pieces in its puzzle thus bringing order where
   chaos was before, and that I have touched greatness and eternity, would make
   me happy beyond words.

  The very inkling of such things potentially happening nearby was a source of
  great excitement for me. The three years I spent with Ne'eman's group shaped
  my attitude for the rest of my career.

  While I do most of the time routine research, I still hope for some
   inspiration and revelation. And I keep missing opportunities, though
   none close to the great opportunities we missed then...

  And I wonder if notwithstanding Yuval's many collaborators over so many
  years and despite all our shortcomings, we were some of the very best he
  ever had.  And I am proud of it.

\begin{center}
* * * * * * * * * * * * * * *
\end{center}

     About a year later on April 26, 2006 Yuval passed away.

During the last weeks preceding this sad and unexpected event, I was in touch
with him and with Andrew Watson, the biographer.
Fighting to the last moment Yuval was trying to convince him to phrase
more clearly the importance of the paper he had with H.~Goldbergon's 1/3 integral
baryon number and I was trying to help as much as I possibly could.

Having discussed this issue with several eminent physicists Andrew finally wrote something very reasonable.

Still I am deeply saddened when in a new, low level, particle physics text  both SU(3) flavor and the quark model are attributed to Gell-Mann only with no mention of either Yuval or Zweig.

But then I console myself thinking that as long as physicists will discuss hadrons, Yuval's name will live---as (I believe) that he is the one who introduced this name.
\end{document}